\documentclass[graybox]{svmult}

\usepackage{type1cm}        
\usepackage{makeidx}         
\usepackage{graphicx}        
\usepackage{multicol}        
\usepackage[bottom]{footmisc}

\usepackage{newtxtext}       %
\usepackage[varvw]{newtxmath}       
\usepackage{eso-pic}
\newcommand\AtPageUpperMyright[1]{\AtPageUpperLeft{%
 \put(\LenToUnit{0.33\paperwidth},\LenToUnit{-1cm}){%
     \parbox{0.8\textwidth}{\raggedleft\fontsize{9}{11}\selectfont #1}}%
 }}%
\newcommand{\conf}[1]{%
\AddToShipoutPictureBG*{%
\AtPageUpperMyright{#1}
}
}

\usepackage{amsmath}

\usepackage{subcaption}
\usepackage[super]{nth}

\usepackage{siunitx}

\newcommand{\etal}{\textit{et al}.~}

\usepackage{acronym}
\acrodef{CFD}[CFD]{Computational Fluid Dynamics}
\acrodef{DG}[DG]{Discontinuous Galerkin}
\acrodef{FV}[FV]{Finite Volume}
\acrodef{FD}[FD]{Finite Difference}
\acrodef{DGSEM}[DGSEM]{Discontinuous Galerkin Spectral Element Method}
\acrodef{DNS}[DNS]{Direct Numerical Simulation}
\acrodef{LES}[LES]{Large Eddy Simulation}
\acrodef{DDES}[DDES]{Delayed Detached Eddy Simulation}
\acrodef{LSB}[LSB]{laminar separation bubble}
\acrodef{RANS}[RANS]{Reynolds-averaged Navier-Stokes}
\acrodef{STG}[STG]{Synthetic Turbulence Generator}
\acrodef{LE}[LE]{leading edge}
\acrodef{TE}[TE]{trailing edge}
\acrodef{PV}[PV]{passage vortex}
\acrodef{TSV}[TSV]{trailing shed vortex}
\acrodef{TKE}[TKE]{turbulent kinetic energy}
\acrodef{PSD}[PSD]{Power Spectral Density}
\acrodef{SVD}[SVD]{Singular Value Decomposition}
\acrodef{PIV}[PIV]{Particle Image Velocimetry}
\acrodef{LPT}[LPT]{low-pressure turbine}
\acrodef{WENO}[WENO]{weighted essentially non-oscillatory}
\acrodef{KH}[KH]{Kelvin-Helmholtz}
\acrodef{LIC}[LIC]{line integral convolution}
\acrodef{MSER}[MSER]{marginal standard error rule}

\acrodef{POD}[POD]{Proper Orthogonal Decomposition}
\acrodef{SPOD}[SPOD]{Spectral Proper Orthogonal Decomposition}
\acrodef{DMD}[DMD]{Dynamic Mode Decomposition}

\newcommand{\xbycax}{x/C_{\mathrm{ax}}}

\newcommand{\figref}[1]{Fig.~\ref{#1}}

\usepackage{array}
\newcolumntype{P}[1]{>{\centering\arraybackslash}p{#1}}  

\usepackage{tikz}

\newcommand{\ld}{\text{\----}}
\DeclareMathOperator{\diag}{diag}

\makeindex             

\begin{document}

\title*{Modal analysis of high-fidelity simulations in turbomachinery}
\author{Christian Morsbach \and Bjoern F. Klose \and Michael Bergmann \and Felix M. Möller}

\institute{
Christian Morsbach \at Institute of Propulsion Technology, German Aerospace Center (DLR), Cologne, Germany, \email{christian.morsbach@dlr.de}
\and
Bjoern F. Klose \at Institute of Test and Simulation for Gas Turbines, German Aerospace Center (DLR), Cologne, Germany, \email{bjoern.klose@dlr.de}
\and
Michael Bergmann \at Institute of Propulsion Technology, German Aerospace Center (DLR), Cologne, Germany, \email{michael.bergmann@dlr.de}
\and
Felix Möller \at Institute of Test and Simulation for Gas Turbines, German Aerospace Center (DLR), Cologne, Germany, \email{felix.moeller@dlr.de}
}
\maketitle
\conf{CUFS-2024-4, Cambridge, United Kingdom, 4-5 March 2024} 

\abstract*{
We revisit recently published high-fidelity implicit large eddy simulation datasets obtained with a high-order discontinuous Galerkin spectral element method and analyse them using Proper Orthogonal Decomposition (POD) as well as Spectral Proper Orthogonal Decomposition (SPOD).
The first configuration is the MTU T161 low-pressure turbine cascade with resolved end wall boundary layers in a version with and without incoming turbulent wakes.
We focus on the behaviour of the laminar separation bubble and the secondary flow system and how these phenomena are affected by incoming wakes.
The second configuration is a transonic compressor cascade in which we analyse the unsteady behaviour of the shock wave boundary layer interaction.
Throughout the paper, we try to discuss not only the flow physics but also insights into how the modal decomposition techniques can help facilitate understanding and where their limitations are.
}

\abstract{
We revisit recently published high-fidelity implicit large eddy simulation datasets obtained with a high-order discontinuous Galerkin spectral element method and analyse them using Proper Orthogonal Decomposition (POD) as well as Spectral Proper Orthogonal Decomposition (SPOD).
The first configuration is the MTU T161 low-pressure turbine cascade with resolved end wall boundary layers in a clean version and one with incoming turbulent wakes.
We focus on the behaviour of the laminar separation bubble and the secondary flow system and how these phenomena are affected by incoming wakes.
The second configuration is a transonic compressor cascade in which we analyse the unsteady behaviour of the shock wave boundary layer interaction.
Throughout the paper, we try to discuss not only the flow physics but also insights into how the modal decomposition techniques can help facilitate understanding and where their limitations are.
}

\section{Introduction}
\ac{CFD} simulations have revolutionized our ability to predict and understand complex flow phenomena.
With increasing computing resources and the development of high-order methods, simulations produce more and more high-fidelity datasets, which are difficult to analyse.
While these datasets have the potential to improve our understanding of flow physics, the sheer amount of data, and its temporal and spatial resolution, make gaining insights a challenging task.
Hence, most \ac{LES} studies only discuss time-averaged quantities or features based on snapshots.

To facilitate the analysis of these datasets, researchers have recently developed data analysis techniques, such as \ac{POD}~\cite{Berkooz1993,Weiss2019}, \ac{SPOD}~\cite{Schmidt2020}, and \ac{DMD}~\cite{Schmid2010}, which have gained significant attention in the computational fluid dynamics community.
These methods allow for the extraction of the most energetic modes (\ac{POD}) and sorting by frequency content (\ac{SPOD}).

Traditionally, these modal decomposition techniques have been applied to experimental data.
Some recent examples include \ac{POD} of \ac{PIV} data from a centrifugal compressor inlet~\cite{Banerjee2022}, \ac{POD} and \ac{DMD} applied to analyse \ac{PIV} data of a laminar separation bubble undergoing transition to turbulence~\cite{Lengani2016,Alessandri2019,Dotto2021}, \ac{POD} used to investigate the interaction of high- and low-pressure turbine based on unsteady total pressure data in planes between the blade rows~\cite{Dellacasagrande2019}, or an investigation of transonic buffet using \ac{DMD} of \ac{PIV} data~\cite{FeldhusenHoffmann2021}.
Very early numerical studies were based on unsteady \ac{RANS} data, e.g.~\cite{Cizmas2003}, but more recently, turbulence resolving approaches such as hybrid \ac{RANS}/\ac{LES} or \ac{LES} have become feasible.
Fiore \etal\cite{Fiore2023} recently applied the \ac{SPOD} method to a low-pressure turbine flow with laminar inflow, as well as with wakes generated by a row of upstream cylinders. 
They found that the cylinder wakes amplify certain modes over the blade and increases turbulence downstream of the trailing edge over a reduced frequency bandwidth.
Mechanisms of \ac{TE} cutback film cooling flows at different blowing rations have been investigated using \ac{SPOD} on \ac{DDES} data by Wang \etal~\cite{Wang2022}.
Here, the authors analysed how the structure of vortex shedding changes with blowing ratio.
He \etal~\cite{He2021} recently presented an extensive analysis of the compressor tip leakage flow using \ac{SPOD} on data obtained with a \ac{DDES}.
They used the complete 3D flow field in the tip region of the blade as input for the \ac{SPOD} and identified low-rank behaviour over a range of frequencies associated with an oscillation of the tip leakage vortex and vortex shedding after its breakdown.

At the German Aerospace Center (DLR), we generate \ac{LES} data with a numerical test rig for turbomachinery based on the high-order \ac{DGSEM}.
In this study, we will reconsider previously published high-fidelity datasets of turbomachinery flows~\cite{Klose2023,Morsbach2023} generated with our \ac{DGSEM} solver TRACE~\cite{Bergmann2023} and extract the dominant dynamic features using \ac{POD} and \ac{SPOD}.
The datasets include the interaction of wakes with downstream low-pressure turbine blades and the shock boundary layer interaction in a compressor cascade.
The study will discuss the advantages and limitations of each method and provide insights into how these techniques can be used to gain a deeper understanding of complex flow phenomena.

\section{Modal decomposition techniques}
In this section, we will very briefly introduce \ac{POD} and \ac{SPOD} to clarify the terminology.

\subsection{Proper Orthogonal Decomposition (POD)}
\ac{POD} by itself is not a new method.
Its theoretical background and formulation have been reviewed decades ago~\cite{Berkooz1993}.
The formulation here is inspired by the review of Schmidt and Colonius~\cite{Schmidt2020} as well as the tutorial by Weiss~\cite{Weiss2019}.
For a consistent derivation of how a snapshot \ac{POD} can be conveniently computed on non-equidistant grids using a \ac{SVD}, we refer the interested reader to the appendix.

We define the matrix of $M$ snapshots of fluctuating quantities (mean subtracted) where each snapshot is a row and spatial dimensions as well as multiple variables are flattened to size $N$:
\begin{equation}  \label{eqn:snapshotMatrix}
\mathbf X = 
\begin{pmatrix}
\ld & \mathbf x_1^H & \ld\\
& \vdots \\
\ld & \mathbf x_M^H & \ld\\
\end{pmatrix}
\in \mathbb C^{M \times N}.
\end{equation}
The \ac{POD} is based on the sample covariance matrix
\begin{equation} \label{eqn:covarianceMatrix}
\mathbf C = \frac{1}{M-1} \mathbf X^H \mathbf X \in \mathbb C^{N \times N}
\end{equation}
through an eigenvalue problem with a weight matrix $\mathbf W$ and inner product $\langle \mathbf x_1, \mathbf x_2 \rangle = \mathbf x_1^H \mathbf W \mathbf x_2$~\cite{Schmidt2020}:
\begin{equation} \label{eqn:POD}
\mathbf C \mathbf W \mathbf \Phi = \mathbf \Phi \mathbf \Lambda,
\quad
\mathbf W = \diag\left(w_1, \dots, w_N\right).
\end{equation}
Here, $\mathbf \Phi \in \mathbb C^{N \times M}$ is the matrix of \ac{POD} modes or eigenvectors and $\mathbf \Lambda = \diag\left(\lambda_1, \dots, \lambda_M\right)$ is the matrix of eigenvalues.
The modes are orthonormal under the scalar product defined above:
\begin{equation} \label{eqn:POD_modes_orthonormal}
    \mathbf \Phi^H \mathbf W \mathbf \Phi = \mathbf 1.
\end{equation}
A linear combination of the modes to recover the original snapshots:
\begin{equation}
    \mathbf X = \mathbf A \mathbf \Phi^T = \sum_j a_{ij} \Phi_{jk} = \sum_j \mathbf {\tilde X}_j
    \in \mathbb R^{M \times N}.
\end{equation}
This can be interpreted as sum of modal contributions with the time coefficients determined by
\begin{equation} \label{eqn:timeCoefficients}
\mathbf A = \mathbf X \mathbf W \mathbf \Phi
\in \mathbb C^{M \times N}.
\end{equation}

\subsection{Spectral Proper Orthogonal Decomposition (SPOD)}

In this section, we briefly review the implementation of the \ac{SPOD} algorithm based on Welch's periodogram method, which is used in this work. 
For a detailed description of the theoretical background, we refer the reader to Towne, Schmidt and collaborators~\cite{Towne2018, Schmidt2020}, as well as to Mengaldo \& Maulik~\cite{Mengaldo2021} for the implementation.

Just as done for the \ac{POD}, we assemble the matrix of snapshots $\mathbf{X}$~\eqref{eqn:snapshotMatrix}.
Next, we apply Welch's spectral estimation method by decomposing the matrix along its time axis into $N_\mathrm{blk}$ overlapping blocks and, assuming that each realisation in the snapshot matrix occurs periodically in time, apply a discrete Fourier transform on each block after multiplication of the data by a Hamming window.
The result is a set of $N_\mathrm{blk}$ Fourier transformed blocks $\widetilde{\mathbf{X}}_k$ with $N_\mathrm{fft}$ frequencies each. 

For each frequency, we assemble the matrix $\widetilde{\mathbf{X}}$
\begin{equation}  \label{eqn:snapshotMatrixTransformed}
\widetilde{\mathbf{X}} = 
\begin{pmatrix}
\ld & \widetilde{\mathbf{x}}_1^H & \ld\\
& \vdots \\
\ld & \widetilde{\mathbf{x}}_{N_\mathrm{blk}}^H & \ld\\
\end{pmatrix}
\in \mathbb C^{N_\mathrm{blk} \times N},
\end{equation}
where, again, $N$ is the number of spatial points times the number of variables and $N_\mathrm{blk}$ is the number of blocks.
For each frequency, we compute the covariance matrix
\begin{equation} \label{eqn:covarianceMatrixTranformed}
\widetilde{\mathbf{C}} = \frac{1}{N_\mathrm{blk}}  \widetilde{\mathbf{X}}^H \widetilde{\mathbf{X}}\in \mathbb C^{N \times N},
\end{equation}
and obtain the \ac{SPOD} modes by solving the eigenvalue decomposition of the covariance matrix for each frequency separately~\cite{Schmidt2020}:
\begin{equation} \label{eqn:SPOD}
\widetilde{\mathbf{C}} \mathbf W \mathbf \Phi = \mathbf \Lambda \mathbf \Phi,
\quad
\mathbf W = \diag\left(w_1, \dots, w_N\right).
\end{equation}
Here, $\mathbf{W}$ is a weight matrix, $\mathbf{\Phi}$ are the eigenvectors and $\mathbf{\Lambda}$ the eigenvalues.
Because the weight matrix is problem specific, we will discuss its form for each application separately.

\section{Application to turbomachinery flows}
In the following, we will apply both \ac{POD} and \ac{SPOD} to three different datasets with different flow phenomena.
We will begin by investigating the transitional flow through the MTU T161 \ac{LPT} with focus on separation-induced transition around midspan and large-scale unsteadiness of the secondary flow system.
Additionally, we will analyse the influence of wakes from an upstream row.
Finally, we will study shock wave boundary layer interaction in a linear compressor cascade.

\subsection{Separation-induced transition and secondary flow effects}

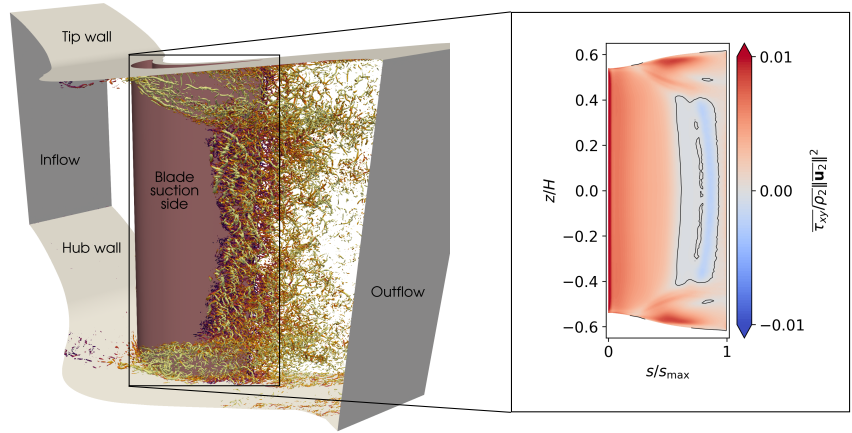
\begin{figure}[t]
    \centering
    \begin{tikzpicture}[]
        \node[rectangle,inner sep=0] at (0,0) {\includegraphics[width=0.5\textwidth]{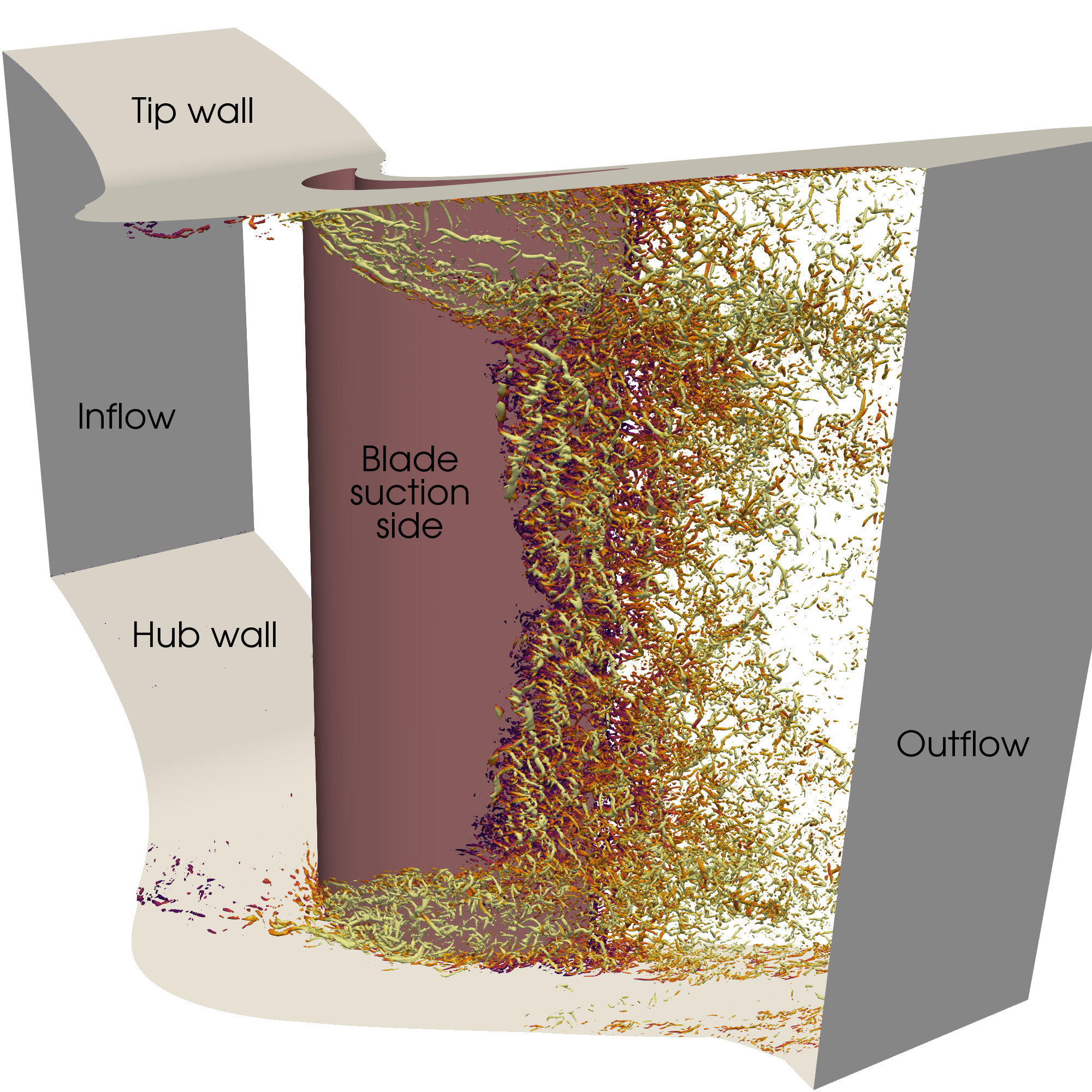}};
        \node[rectangle,draw=black,inner sep=0.2cm] (mean) at (6,0) {\includegraphics[width=0.35\textwidth]{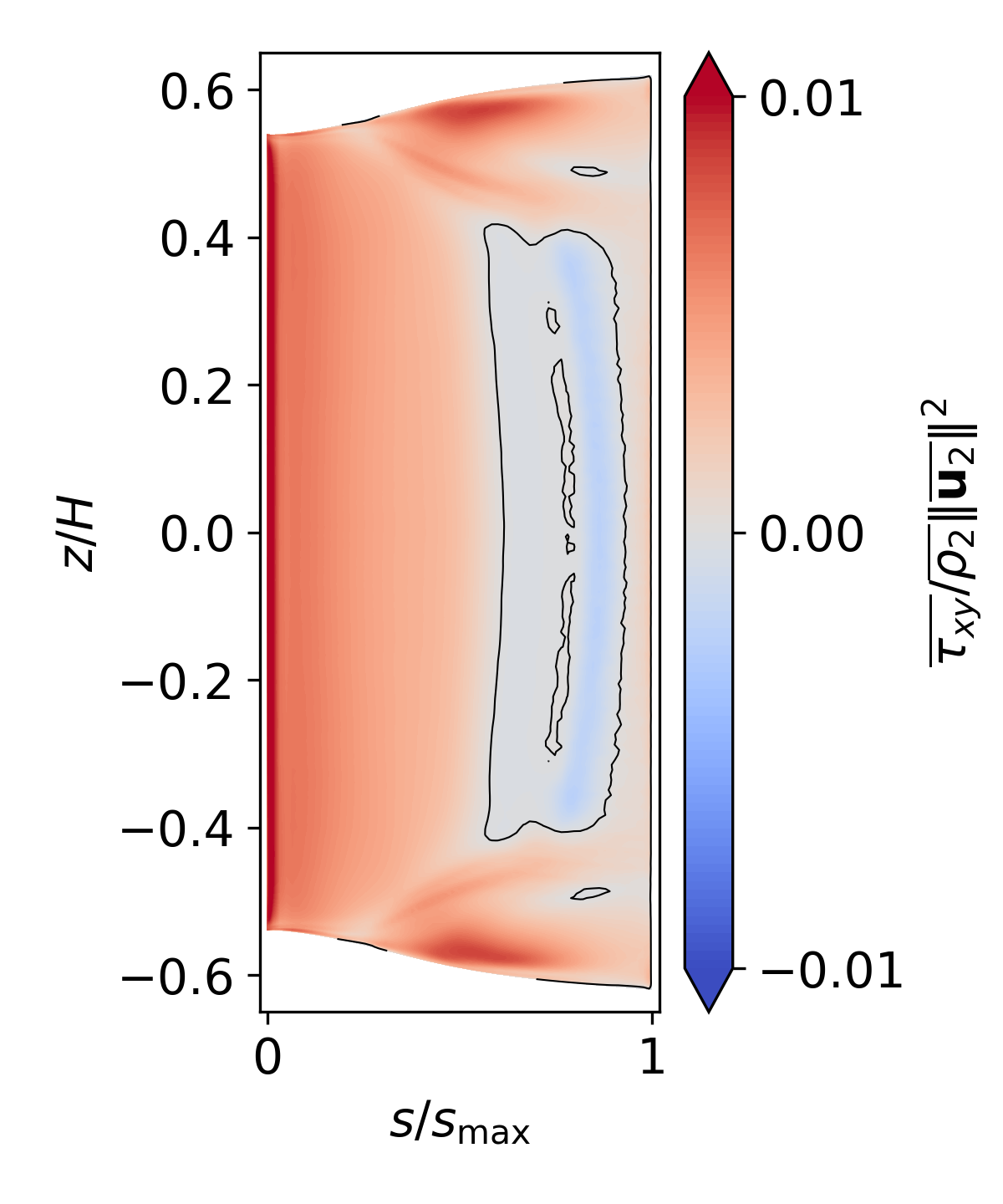}};
        \node[rectangle,draw=black,minimum width = 2cm, 
    minimum height = 4.4cm] (zoom) at (-0.33, -0.11) {};
        \draw [] (zoom.north west) -- (mean.north west);
        \draw [] (zoom.south west) -- (mean.south west);
    \end{tikzpicture}
    \caption{Illustration of MTU T161 simulation domain with instantaneous vortex structures and mean in-plane wall shear stress $\tau_{xy}$ on the blade suction side with $\tau_{xy} = 0$ as black contour line}
    \label{fig:T161:mean}
\end{figure}

We analyse the turbulent flow through an \ac{LPT} cascade at an off-design Reynolds number of $90\,000$ and a Mach number of 0.65 computed as implicit \ac{LES} with a high-order \ac{DGSEM} scheme~\cite{Morsbach2023}.
We use a dataset containing the time-resolved wall shear stress vector on the blade suction side spanning 100 convective time units $t_\mathrm{c}$ based on chord length and outlet velocity with 3668 samples.
The original sampling was 366 samples per $t_\mathrm{c}$ but we use only a tenth of that for the modal analysis, since that already covers the frequency range of interest.
\figref{fig:T161:mean} shows the configuration with a view of the blade suction side of the configuration with instantaneous vortex structures visualised by the Q-criterion to illustrate the separation induced transition around midspan and the secondary flow regions close to the hub and tip walls.
On the right-hand side, the mean in-plane wall shear stress $\tau_{xy}$ on the blade suction side surface over the arc length from the \ac{LE} at $s=0$ to the \ac{TE} at $s / s_\mathrm{max} = 1$ is presented in the way we will display the mode shapes throughout this section.
The $z$ coordinate is normalised by the upstream channel height $H$ allowing to see the diverging end walls.
We will employ classical Fourier analysis, space-only \ac{POD} and \ac{SPOD} to investigate a potential connection between the \ac{KH} based transition mechanism around midspan and oscillations of the corner flow.

\begin{figure}[t]
    \centering
    \includegraphics[width=\textwidth]{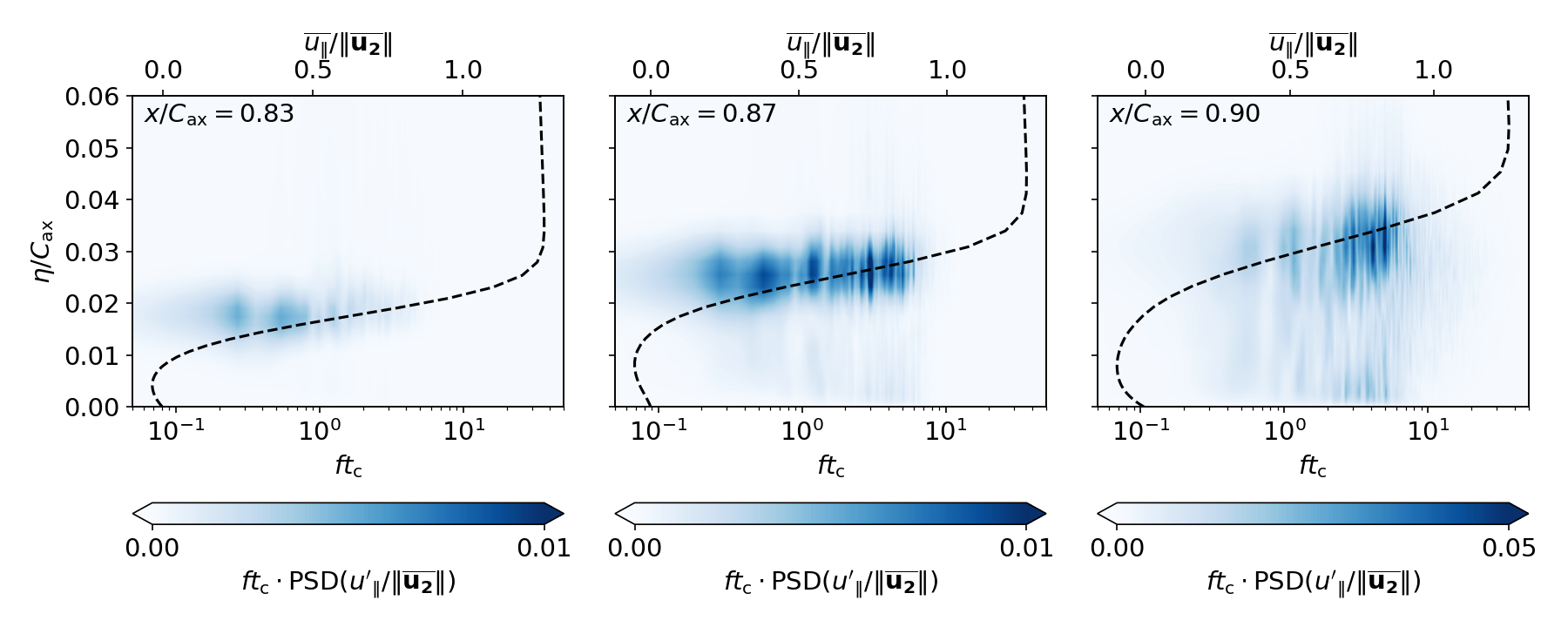}
    \caption{Midspan premultiplied spectrum of $u'_\Vert$ for different distances $\eta$ from the solid wall in boundary layer cuts of the MTU T161 blade suction side at $\xbycax = 0.83$ (pre-transitional), $\xbycax = 0.87$ (transition onset) and $\xbycax = 0.90$ (half way to fully turbulent)}
    \label{fig:T161:midspan_spectrum}
\end{figure}

In addition to the surface dataset, we use three wall-normal cuts through the suction side boundary layer at midspan to be able to relate frequencies to physical mechanisms in the flow.
\figref{fig:T161:midspan_spectrum} shows premultiplied frequency spectra of the wall-parallel fluctuating velocity $u'_\Vert$ for these cuts varying with the distance from the solid wall.
For orientation, the mean velocity profiles are plotted as black dashed lines.
In the cut at $\xbycax = 0.83$, where the separated shear layer is still laminar, only non-dimensional frequencies lower than 1 (time scales larger than a convective time unit) show noticeable content in the region of strongest shearing.
Further downstream at $\xbycax = 0.87$, where the flow begins its transition to turbulence, non-dimensional frequencies in the range 3-5 become excited due to the \ac{KH} instability and have already surpassed the intensity found in the lower frequencies.
These instabilities can be found to grow even further at $\xbycax = 0.90$, where the maximum \ac{TKE} in the boundary layer is at 50\% of the maximum, reached slightly downstream.
At this position, they dominate the whole spectrum and peaks at these frequencies can not only be found in the shear layer but also close to the solid wall.
Observe the changed scale of the colourbar.
We will find both the low frequency and high frequency effects in the modal analysis below.

\begin{figure}[t]
    \sidecaption[t]
    \includegraphics[width=0.49\textwidth]{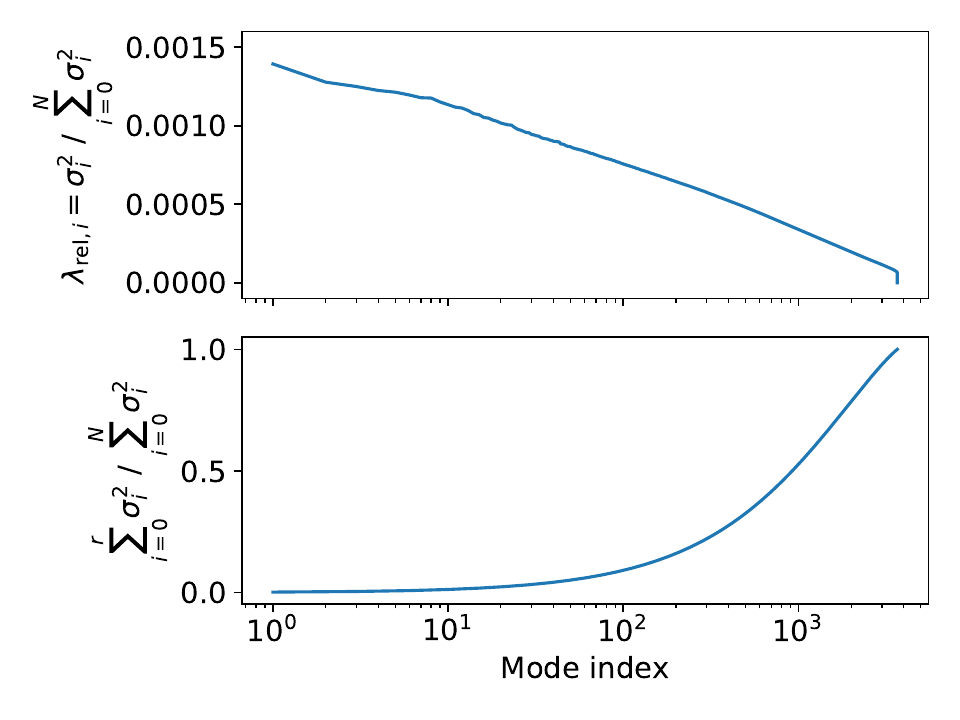}
    \caption{POD eigenvalues of MTU T161 suction side wall shear stress (\textit{top}) and cumulative eigenvalues (\textit{bottom}); values for Gaussian random data given as indication of noise threshold}
    \label{fig:T161:POD_eigenvalues}
\end{figure}

Since our dataset has points clustered towards the leading and trailing edge, as well as towards the end walls, we use the element area as weights to perform a \ac{POD} using an \ac{SVD} with \eqref{eqn:SVD}, \eqref{eqn:SVD_to_POD_eigenvalues} and \eqref{eqn:SVD_to_POD_modes}.
The input variables are the two independent components of the fluctuating wall shear stress vector $\tau_{xy}'$ and $\tau_z'$, equally weighted using the weighted 2-norm~\cite{Schmidt2020}:
\begin{equation}
    \mathbf X = \left[\tau_{xy}',\; \tau_z'\right]^\mathrm{T}, \quad \mathbf W = \int_V\diag\left(\mathrm{d}V,\; \mathrm{d}V\right).
\end{equation}

The resulting eigenvalues (\textit{top}) and the cumulative energy contained in the first $r$ modes (\textit{bottom}) are shown in \figref{fig:T161:POD_eigenvalues}.
It can be instantly noted that, in contrast to many textbook examples of \ac{POD}, there is no large-scale organisation in the flow field responsible for a great portion the fluctuating energy.
On the contrary, the most energetic mode only accounts for 0.14\% while the tenth mode still accounts for 0.11\%.
Yet, the modes stand out above the noise threshold and the mode shapes offer some insights into the flow physics.

\begin{figure}[t]
    \centering
    \includegraphics[width=\textwidth]{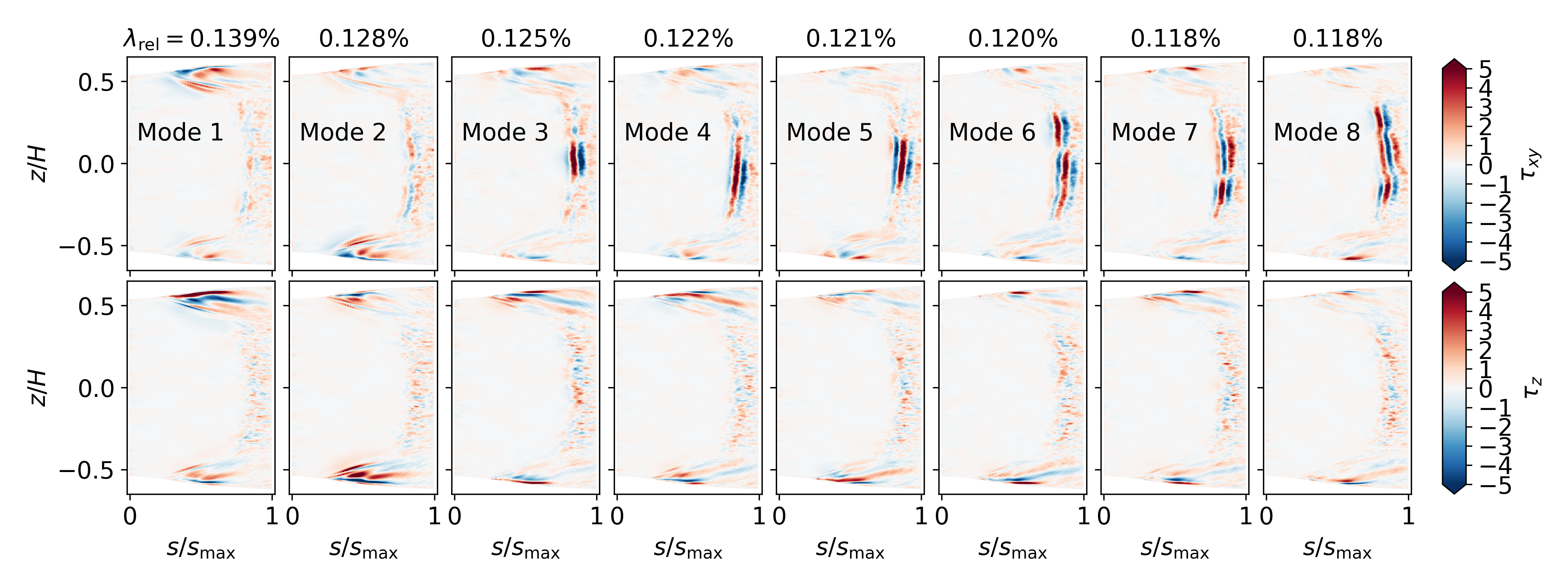}
    \caption{MTU T161: First eight \ac{POD} modes for in-plane (\textit{top}) and spanwise (\textit{bottom}) wall shear stress $\tau_\mathrm{xy}, \tau_\mathrm{z}$}
    \label{fig:T161:POD_modes}
\end{figure}

\figref{fig:T161:POD_modes} shows the first eight \ac{POD} modes.
Note that the modes are orthonormal by definition, so the corresponding eigenvalues are given in the title to indicate their relative strengths.
The first two modes exhibit structures close to the end walls while the area around midspan shows relatively low amplitudes and rather noisy data.
Mode 1 has larger amplitudes at the tip wall (positive $z$) in both components of the wall shear stress while mode 2 shows a very similar structure basically mirrored at the $z=0$ plane.
This corresponds to a periodic displacement of flow from the end walls into the channel and can be associated with an oscillation of the \ac{PV}.
The third mode is the first one to be dominated by structures associated with the \ac{KH} instability in the separated shear layer around midspan.
It is notable that the large amplitudes are only found close to midspan and are perfectly aligned with the $z$ axis.
Larger structures close to the end walls can still be seen here, yet at a lower amplitude compared to modes 1 and 2.
A spanwise variation of the \ac{KH} structures can be found in the following modes be it either in the form of a tilt (mode 4, 5, 8) or an inversion of the phase at some spanwise location (mode 6 and 7).
All the modes featuring \ac{KH} structures still show some amplitude at the end walls suggesting that these phenomena might be connected.

\begin{figure}[t]
    \centering
    \mbox{
    \includegraphics[width=0.49\textwidth]{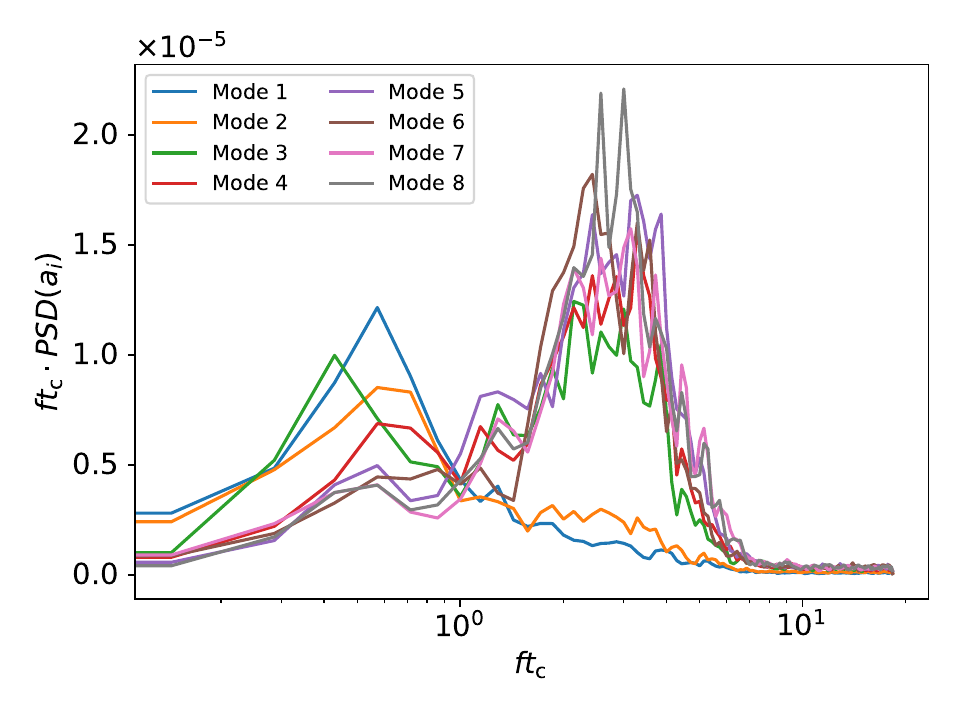}
    \hfill
    \includegraphics[width=0.49\textwidth]{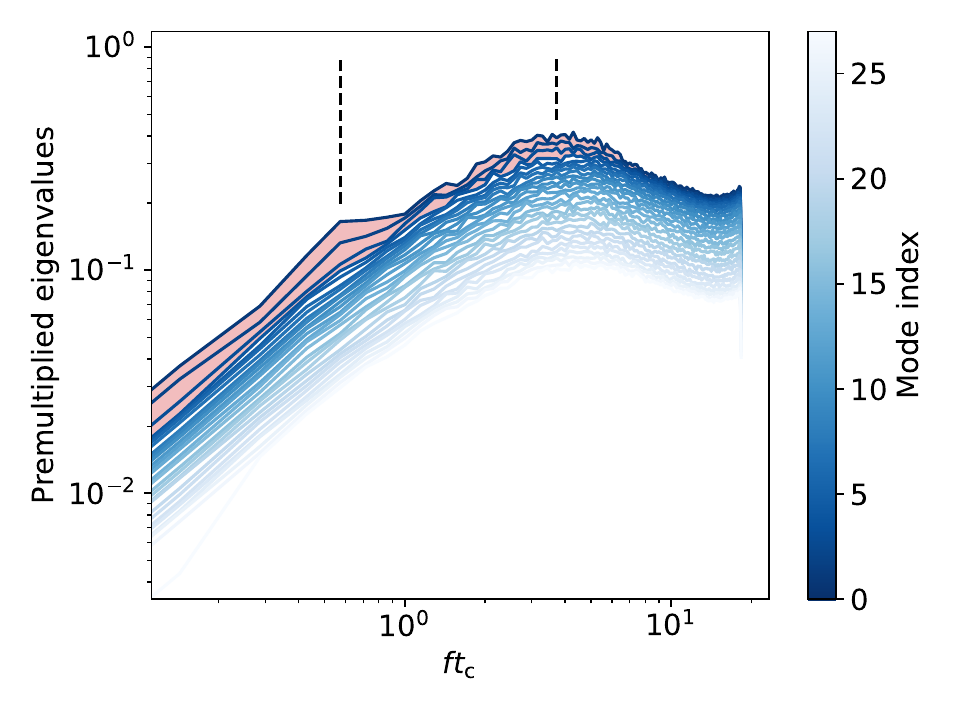}
    }\\    
    \mbox{
    \makebox[0.49\textwidth][c]{(a) Premultiplied PSD of \ac{POD} time coefficients}
    \makebox[0.49\textwidth][c]{(b) Premultiplied \ac{SPOD} eigenvalues}
    }
    \caption{Spectra for the MTU T161 without incoming wakes}
    \label{fig:T161:spectrum_POD_timeCoefficients}
\end{figure}

A look at the \ac{POD} time coefficients, especially at their spectral content, can reveal additional information on the modes.
\figref{fig:T161:spectrum_POD_timeCoefficients} (a) shows the premultiplied \ac{PSD} of the time coefficients for the first eight modes, obtained via \eqref{eqn:timeCoefficients}, computed with Welch's method with 256 samples per segment.
Two distinct features can be identified: a low-frequency peak around 0.5 and a high frequency peak around 3.
The first two modes, which do not show any sign of \ac{KH} structures, clearly peak at the low frequency but show no contribution in the high frequency range.
On the other hand, all modes with \ac{KH} structures have the most significant contribution to the spectrum in this high frequency peak.
While most of these modes do not show contributions above the noise level in the low frequency range, mode 3 is an exception: it features a low frequency peak in the same order of magnitude as modes 1 and 2.
This suggests some connection of the oscillation of the secondary flow system with the \ac{KH} shedding through this mode shape.

\begin{figure}[t]
    \centering
    \includegraphics[width=\textwidth]{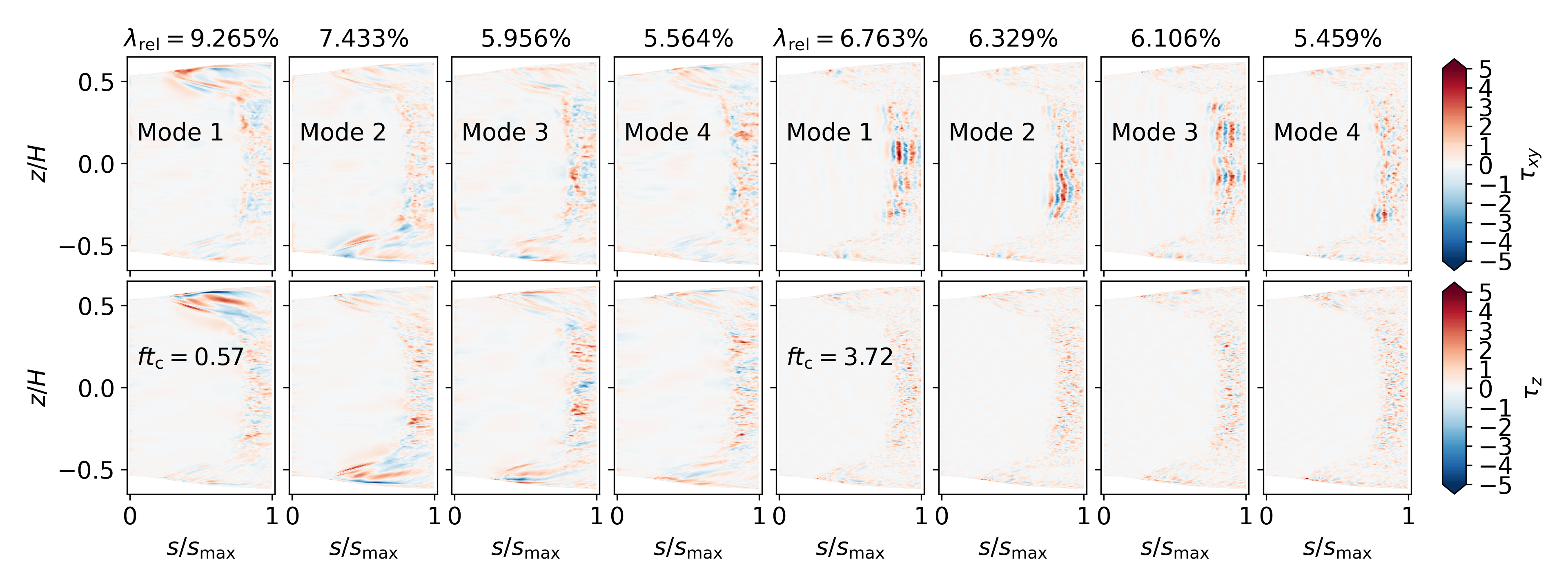}
    \caption{MTU T161: First 4 \ac{SPOD} modes for in-plane (\textit{top}) and spanwise (\textit{bottom}) wall shear stress $\tau_\mathrm{xy}, \tau_\mathrm{z}$ at $ft_\mathrm{c} = 0.57$ and $3.71$}
    \label{fig:T161:SPOD_modes}
\end{figure}

At this point, it can be helpful to resort to \ac{SPOD} to be able to obtain optimal spatial modes corresponding to a single frequency.
We use exactly the same dataset with weights chosen to represent a weighted 2-norm~\cite{Schmidt2020}, a bin size of 256 and an overlap of 50\%.
\figref{fig:T161:spectrum_POD_timeCoefficients} (b) shows the premultiplied \ac{SPOD} spectrum  with the area between mode 1 and 4 shaded in red to indicate the energy contained within the first three modes.
The spectrum peaks in the same high frequency range as the \ac{POD} time coefficient spectra discussed above.
As expected from the above discussion of the \ac{POD}, the separation between the first and second mode is not as clear as in many textbook examples.
Nevertheless, the first three to four modes stand out in terms of separation between the modes and suggest that some kind of low-rank behaviour can possibly be identified~\cite{He2021}.
We, therefore, examine the first four modes at the frequencies 0.57 and 3.72 in \figref{fig:T161:SPOD_modes}.
The choice of former is due to the subtle low-frequency peak while the latter is roughly the maximum premultiplied eigenvalue over all frequencies.
Note, that slight changes to the investigated frequency do not qualitatively change the mode shapes in this case.

The first four columns show the first four modes of $ft_\mathrm{c} = 0.57$.
To present an indication for the energy of the respective mode, the mode eigenvalue $\lambda_\mathrm{rel}$ normalised with the sum over all eigenvalues at that frequency is given in the column header.
The first two \ac{SPOD} modes at $ft_\mathrm{c} = 0.57$ show a great resemblance with the first two \ac{POD} modes, which is a consistent picture since the frequency analysis of the latter revealed that most of their energy is found at this low frequency.
Looking at the higher modes, there is again some indication that the oscillation of the secondary flow system has an effect on the midspan flow, although the data in the separation bubble is rather noisy.
In contrast to our argument above using the \ac{POD} analysis, however, we see no direct connection with \ac{KH} structures but a larger scale oscillation in the third mode.
\ac{KH} structures in the \ac{LSB}, on the other hand can be clearly identified in the in-plane wall shear stress for all modes at $ft_\mathrm{c} = 3.71$.
Similar to what was found above in the higher \ac{POD} modes, the first mode at this frequency shows a vertical alignment of the \ac{KH} structures while the higher modes introduce some tilting.
The spanwise wall shear stress shows mainly turbulent noise.
An interesting difference to the \ac{POD} concerns the connection of the \ac{KH} structures and the oscillation of the secondary flow system:
while the \ac{POD} suggested some combined mechanism through mode 3, the \ac{SPOD} suggests that the two phenomena are, indeed, not directly coupled as no significant structures in the secondary flow region can be found at $ft_\mathrm{c} = 3.71$.
This underlines the importance of \ac{SPOD} to avoid drawing false conclusions due to the missing frequency resolution of \ac{POD}.
It has to be mentioned, though, that the \ac{SPOD} modes are rather noisy and for a more quantitative analysis, more samples in excess of the available 100 convective time units would be beneficial.
Very recently, a sub-sampling method was suggested to overcome the problems associated with short time signals by sacrificing the high-frequency part of the spectrum to be able to obtain more realisations of low-frequency phenomena~\cite{Schneider2023}.

\subsection{Blade row interaction}

Besides the clean inflow configuration with freestream turbulence only, a dataset with incoming periodic disturbances generated by cylindrical wake generators was presented in~\cite{Morsbach2023}.
\ac{POD} was only used to obtain less noisy phase averages in the wake plane in this paper.
In the following, we perform an \ac{SPOD} analysis consistent with the one above.
The dataset consists of 3009 samples, spanning 82 convective time units corresponding to 50 bar passes.
Again, we use only every tenth sample of the original dataset sampled at 368 per $t_\mathrm{c}$.
The periodic wakes lead to an intermittently separated flow around midspan, which does not show any backflow on average~\cite{Morsbach2023}.

\begin{figure}[t]
    \sidecaption[t]
    \includegraphics[width=0.49\textwidth]{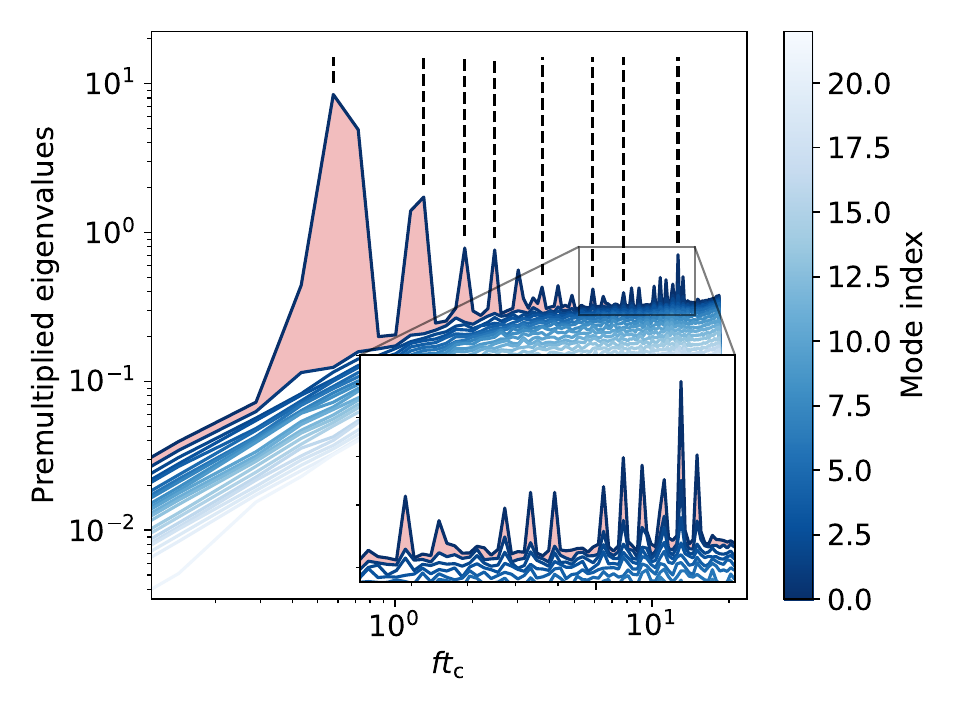}
    \caption{MTU T161: \ac{SPOD} spectrum for the case with incoming wakes; frequencies for mode analysis marked with dashed lines}
    \label{fig:T161:wakes:SPOD_spectrum}
\end{figure}

The \ac{SPOD} spectrum in \figref{fig:T161:wakes:SPOD_spectrum} is dominated by the wake passing frequency $ft_\mathrm{c} = 0.61$ and its harmonics.
Coincidentally, this frequency is very close to the one of the large-scale secondary flow oscillations found in the case without wakes.
The higher harmonics are in a frequency range corresponding to the \ac{KH} shedding.
In the high frequency range $ft_\mathrm{c} > 5$, another set of peaks standing out of the turbulent noise can be found which does not occur in the case without wakes.
In the following, we will analyse the first modes at the frequencies marked with dashed lines.
An inspection of the spectrum and the second modes revealed that these do not generally show larger scale structures with a significant signal to noise ratio.

\begin{figure}[t]
    \centering
    \includegraphics[width=\textwidth]{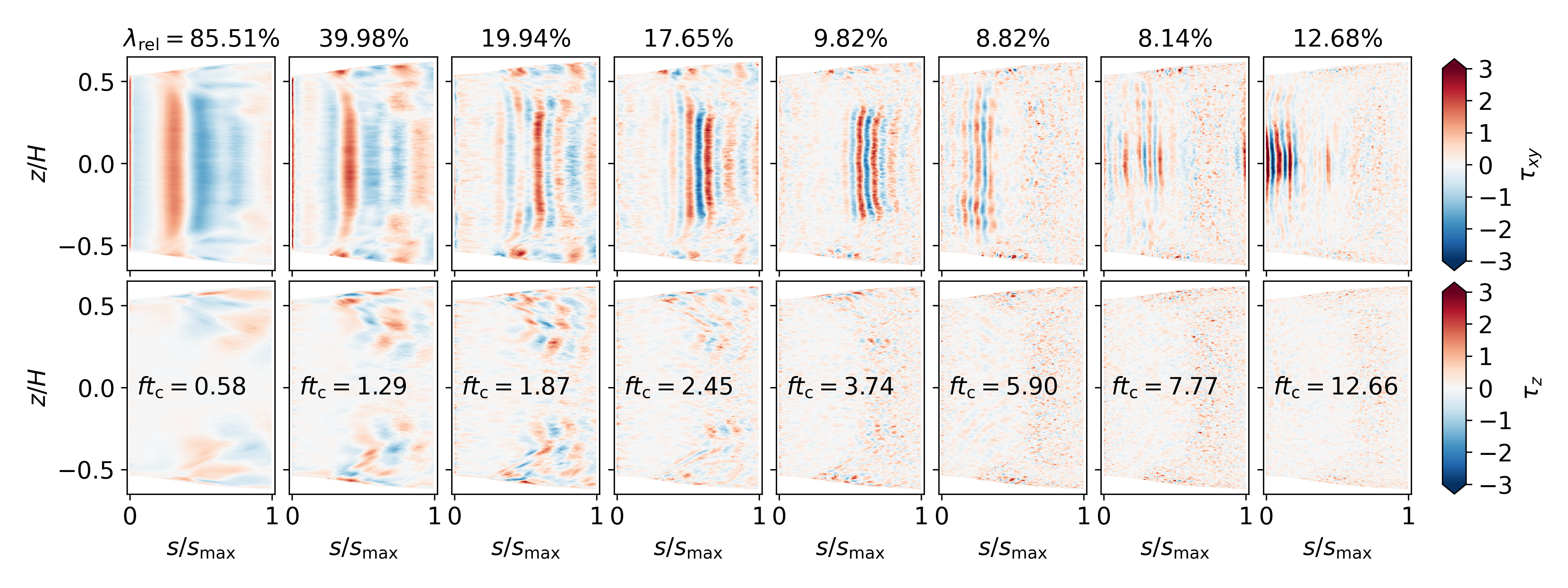}
    \caption{MTU T161: First \ac{SPOD} modes for in-plane (\textit{top}) and spanwise (\textit{bottom}) wall shear stress $\tau_\mathrm{xy}, \tau_\mathrm{z}$ at dominant frequencies}
    \label{fig:T161:wakes:SPOD_modes}
\end{figure}

\figref{fig:T161:wakes:SPOD_modes} shows these modes, again along with their relative eigenvalue normalised with the sum of all eigenvalues at that frequency.
Note that the ratio of the eigenvalues between different frequencies cannot be derived from these but must be obtained from \figref{fig:T161:wakes:SPOD_spectrum}.
The first mode at $ft_\mathrm{c} = 0.58$ is clearly associated with the large-scale wake passing and is responsible for most of the energy at that frequency with mode 2 at $\lambda_\mathrm{rel} = 1.26 \%$ only (not shown in the figure).
Note that due to the choice of blocks and the time step size, the available frequencies in the \ac{SPOD} only approximately match the bar passing frequency and its harmonics.
The large structures in the in-plane wall shear stress describe the periodic separation and reattachment, although this conclusion can only be drawn in combination with knowledge about the local mean values.
In contrast to the case without wakes, the oscillation of the corner separation appears to be dominated by the wake passing as seen in large coherent structures in the both wall shear stress components.
The first three harmonics quickly decrease in eigenvalue and show refined structures corresponding to the same phenomenon.
At frequencies in the range of the \ac{KH} instability of the case without wakes, i.e. $ft_\mathrm{c}$ around 3, the characteristic vertical bands appear in the aft section of the blade.
Again, from this frequency on, at best, a very subtle interaction with the secondary flow region can be identified.

When the frequency is further increased, another set of vertical structures with shorter wavelength can be found upstream, starting near the \ac{LE} at $ft_\mathrm{c} = 12.66$.
The instantaneous flow field in a midspan plane (not shown) revealed, that these are the footprints of upstream traveling acoustic waves originating from the strong vortex shedding of the moving cylinder and their reflections from the adjacent blade pressure side.
The high frequency structures near the \ac{LE} could only be identified using the \ac{SPOD} and cannot be observed in any of the first 30 \ac{POD} modes.
This, again, highlights the usefulness of the \ac{SPOD} in distinguishing between different features of the flow which occur at different frequencies and are possibly hidden within high-energy low-frequency behaviour.

\subsection{Shock wave boundary layer interaction} \label{sec:transonic}
In this section, we analyse the modal content of the flow over a compressor cascade at transonic operating condition. 
The flow is simulated using a \nth{4} order accurate \ac{LES} based on a high-order \ac{DGSEM} scheme.
For a detailed description of the scheme, we refer to Bergmann \etal \cite{Bergmann2023} and references therein.
The linear cascade is computed based on a spanwise domain extrusion of 5\% chord length, where periodic boundary conditions are applied along the pitchwise and the spanwise faces to approximate the flow over an infinite cascade of infinite blades.
The computational domain consists of $2\,276\,829$ hexahedral elements clustered around the blade and the shock location, adding to a total 145.7 million degrees of freedom per equation.
The simulation is evaluated over 21 convective time units ($t_\mathrm{c}$) based on the chord length $c$ and inflow velocity $U_{0}$ and sampled at a non-dimensional frequency of 28 samples per time unit.
We refer the interested reader to Klose \etal ~\cite{Klose2023} for a more detailed description of the setup and discussion of the flow field for this testcase.

Given that the compressor cascade is at transonic operating conditions, the \ac{SPOD} is computed based on the compressible energy norm \cite{Schmidt2020} with input vector 
\begin{equation} \label{eq:input_compr_energy_norm}
    \mathbf X = \left[\rho',\; u_\parallel',\; u_\perp',\; u_z',\; T'\right]^\mathrm{T},
\end{equation}
and weights
\begin{equation} \label{eq:weights_compr_energy_norm}
    \mathbf W = \int_V\diag\left(\frac{\overline{T}}{\gamma\overline{\rho}M_0^2},\; \overline{\rho},\; \overline{\rho},\; \overline{\rho},\; \frac{\overline{\rho}}{\gamma(\gamma-1)\overline{T}M_0^2}\right)\mathrm{d}V,
\end{equation}
where  $\rho$ is the density, $T$ the temperature, $u_i$ the velocity components and $M$ the Mach number. 
All physical quantities here are non-dimensionalised with respect to reference quantities at the inlet measurement plane (indicated by subscript $0$).
A total of 568 snapshots are evaluated for the modal analysis for this testcase, where the bin size for the \ac{SPOD} is 142 snapshots with 50\% overlap. 

\begin{figure}[t]
    \centering
    \mbox{
    \includegraphics[width=0.45\textwidth]{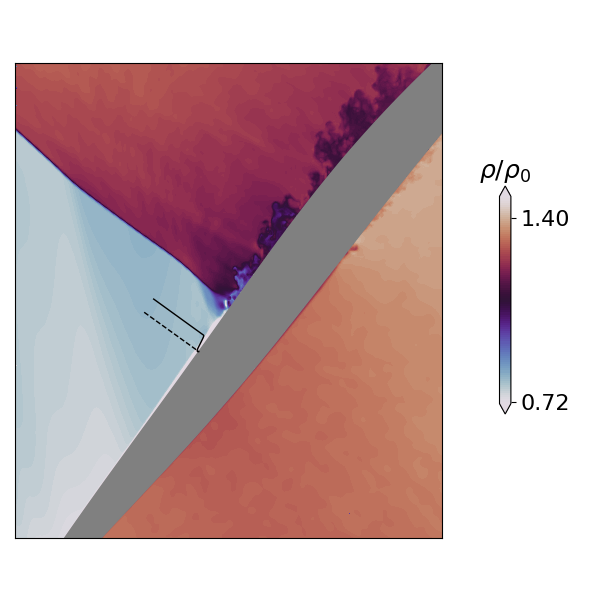}
    \hfill
    \includegraphics[width=0.53\textwidth]{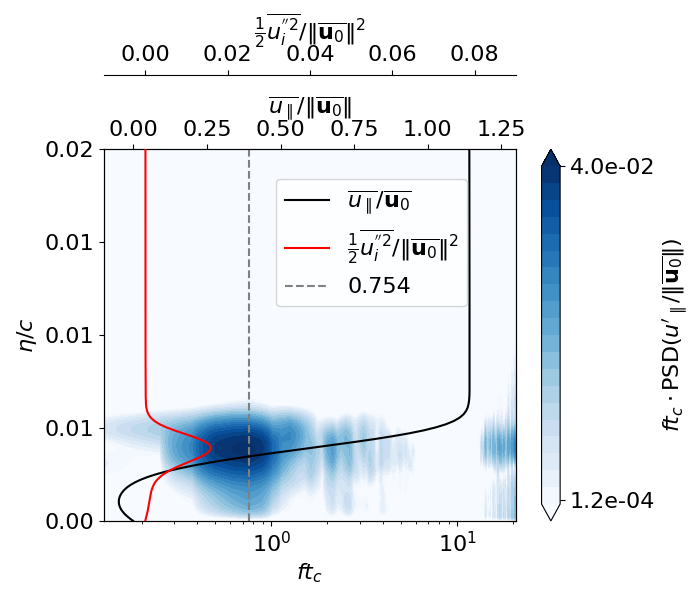}
    }\\
    \mbox{
    \makebox[0.45\textwidth][c]{(a) Density contours}
    \hfill
    \makebox[0.53\textwidth][c]{(b) \ac{PSD} along boundary layer probe}
    }
    \caption{DLR compressor cascade: (a) instantaneous density contours with boundary layer profile indicator in black; (b) contours of the frequency spectrum of $u_\parallel$ along the boundary layer with mean velocity and \ac{TKE} profiles}
    \label{fig:lht:density_psd}
\end{figure}

The flow is characterized by an upstream laminar boundary layer, shock-induced separation and transition to turbulence.
Features of the instantaneous flow field are shown in \figref{fig:lht:density_psd} (a), where contours of the density are shown at the mid-chord section of the blade around the shockwave and the premultiplied frequency spectrum of the wall-parallel velocity component $u_\parallel$ along a wall-normal probe is given in \figref{fig:lht:density_psd} (b).
Additionally, the mean velocity and \ac{TKE}  profiles are indicated, as well as the peak in the spectrum at $f t_\mathrm{c}$ = 0.75.
As shown by Klose \etal~\cite{Klose2023}, low-amplitude oscillations of the normal shock occur in the range $0.6 < f t_\mathrm{c} < 0.8$, as well as high-frequency fluctuations at $f t_\mathrm{c}$ = 2.6, which were associated with a wake mode. 

\begin{figure}[t]
    \centering
    \mbox{
    \includegraphics[width=0.45\textwidth]{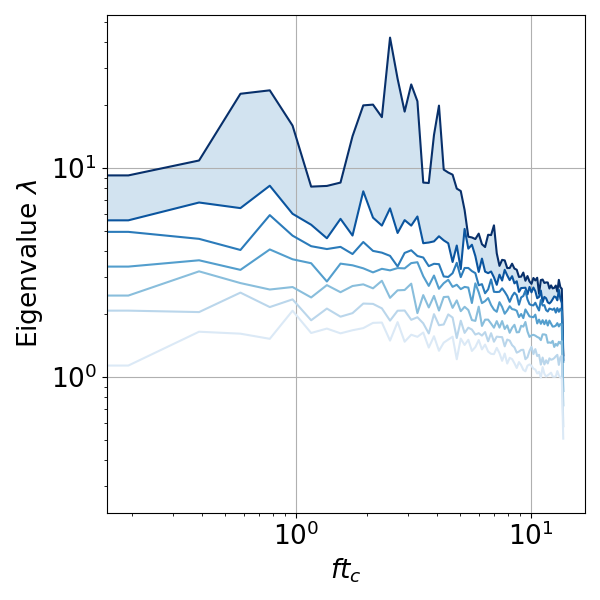}
    \hspace{1cm}
    \includegraphics[width=0.45\textwidth]{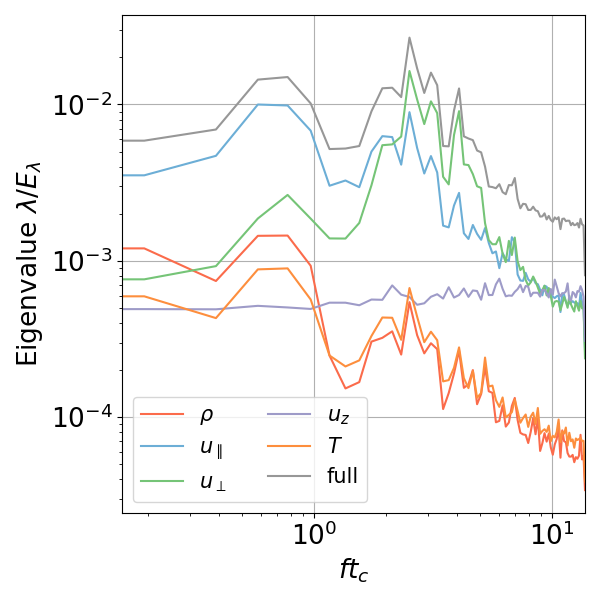}
    }\\
    \mbox{
    \makebox[0.45\textwidth][c]{(a) Eigenvalue spectrum of $q$}
    \hspace{1cm}
    \makebox[0.45\textwidth][c]{(b) Relative contributions}
    }
    \caption{DLR compressor cascade: (a) eigenvalue spectrum of  $q=[\rho, u_\parallel, u_\perp, u_z, T]$; (b) relative contributions to the first mode of the individual components of $q$}
    \label{fig:lht:spod_spectrum}
\end{figure}

The eigenvalue spectrum in \figref{fig:lht:spod_spectrum}(a) is obtained by computing the \ac{SPOD} of the input vector \eqref{eq:input_compr_energy_norm} and the weights \eqref{eq:weights_compr_energy_norm} as a whole, where the shaded area highlights the separation between the first and second \ac{SPOD} modes. 
A distinctly separated peak occurs at $f t_\mathrm{c}$ = 0.77 and a series of separated peaks at around $f t_\mathrm{c}$ = 2.52, closely matching the frequency peak along the boundary layer probe.
To extract the contributions of each term to the overall fluctuating energy, the \ac{SPOD} eigenvalues of the first mode for the individual components (with their respective weights) are shown in \figref{fig:lht:spod_spectrum}(b), together with the result of the \emph{full} input vector.
While the wall-parallel velocity component $u_\parallel'$ dominates the energy in the lower frequency regime, the wall-normal component $u_\perp'$ is the main contributor for the higher frequency peaks.
The components measuring the compressibility effects ($\rho'$, $T'$) also predominantly contribute energy to the low-frequency modes, where they still only make up around 15\% of the total energy.

\begin{figure}[t]
    \centering
    \includegraphics[trim={1cm, 0, 1cm, 0}, clip, width=\textwidth]{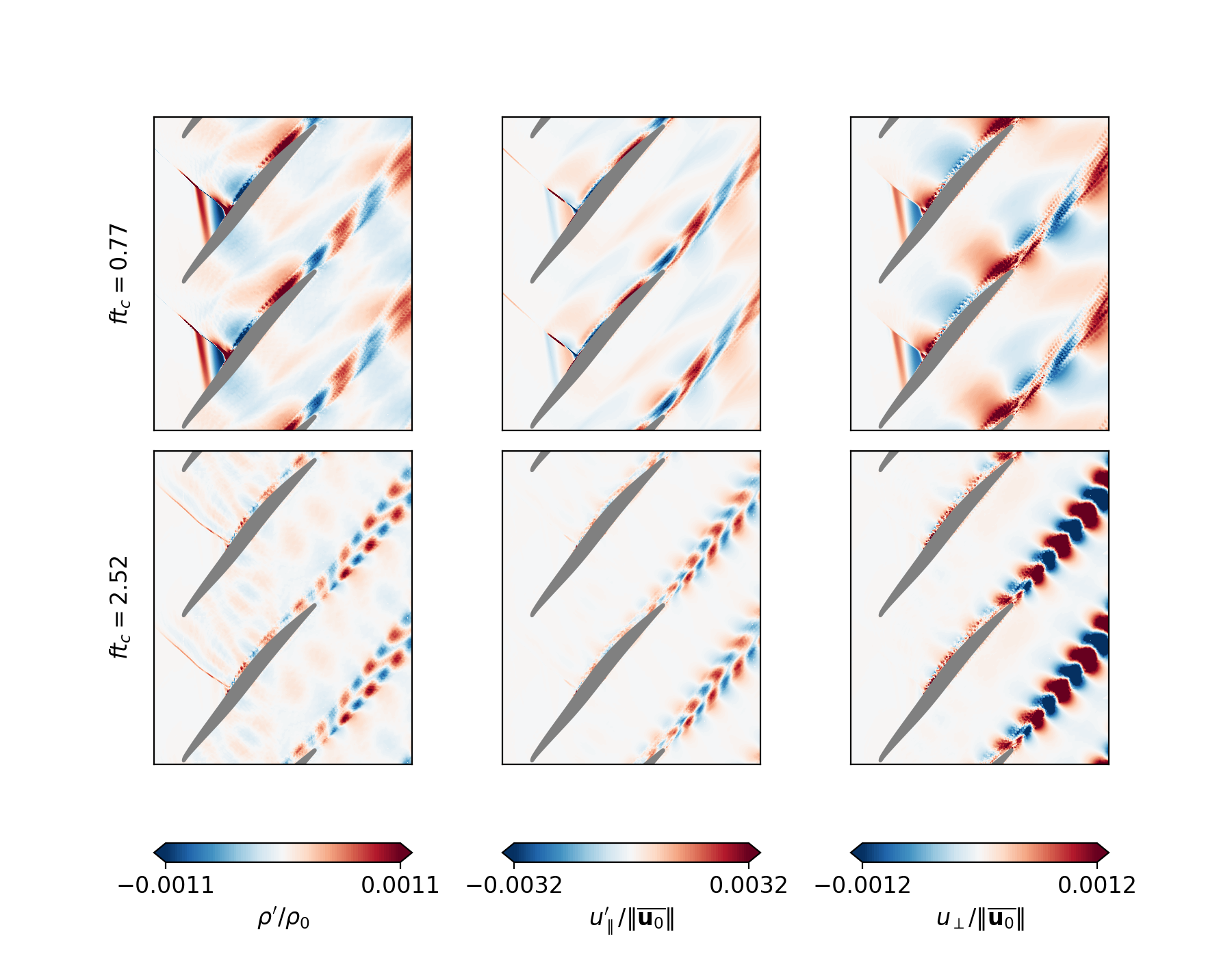}
    \caption{DLR compressor cascade: First SPOD modes of the density (\textit{left}), blade-parallel velocity (\textit{middle}) and blade-normal velocity (\textit{right})}
    \label{fig:lht:spod}
\end{figure}

The shape of the first \ac{SPOD} mode for $f t_\mathrm{c}$ = 0.77 and $f t_\mathrm{c}$ = 2.52 for $\rho'$, $u_\parallel'$ and $u_\perp'$ is given in \figref{fig:lht:spod}. 
At $f t_\mathrm{c}$ = 0.77, the compression area upstream of the normal shock is distinctly highlighted by the density, where the thickening of the boundary layer is accompanied by a mild oblique compression wave.
The movement of the normal shock itself is also tracked by the $u_\parallel'$ velocity component and shows as a distinct, sharp mode structure normal to the blade surface.
At the higher frequency of $f t_\mathrm{c}$ = 2.52, the shock wave is no longer discernible in the velocity components. 
Here, $u_\perp'$ is dominant and its mode shape proves that the oscillating energy is almost exclusively contained in the wake.

\section{Conclusion}
We have presented the application of two modal decomposition techniques, namely \ac{POD} and \ac{SPOD}, to previously published \ac{LES} datasets of turbomachinery flows computed with our high-order \ac{DGSEM} solver TRACE.
While \ac{POD} could be used to draw first conclusions even when large-scale effects did not dominate the turbulent spectrum in energy, the frequency resolution capabilities of \ac{SPOD} allowed for a clearer analysis und separation of effects.
In the \ac{LPT} flow without incoming wakes, the weak but dominant feature was a low-frequency oscillation of the secondary flow system at time scales longer than a convective time unit.
The \ac{SPOD} showed that these oscillations were coupled with a large-scale motion of the \ac{LSB}.
We found, that the \ac{KH} instability, occurring at higher frequencies was easy to detect and did not directly interact with the secondary flow system.
An \ac{SPOD} of the flow through the \ac{LPT} with incoming wakes revealed the wakes' influence on the blade including the secondary flow system while the method was also able to detect the local \ac{KH} mechanism in the intermittent separation bubble.
It also revealed the footprint of very high frequency acoustic perturbations, which originate from the vortex shedding of the upstream moving cylinder.
In this case, "the frequency decomposition benefitted the analysis to reveal low-energy effects otherwise obscured by the dominant large-scale motion.
Applied to the flow around a transonic compressor cascade, \ac{SPOD} was able to discern two effects:
first, low-frequency oscillations are associated with shock-wave boundary layer interaction and show higher contributions of compressible components to the total energy.
Second, high-frequency oscillations are associated with vortex shedding in the wake in which compressibility does not play a significant role.
It has to be noted, that many statements made in this paper depended on a combined analysis of modal decomposition techniques with other methods such as averages, frequency analysis, space-time diagrams or visualisation of the instantaneous flow field.
In this respect, modal decomposition techniques are a valuable addition in the toolbox of unsteady flow analysis in turbomachinery flows.

\bibliographystyle{spmpsci}
\bibliography{literature}

\section*{Appendix}
\addcontentsline{toc}{section}{Appendix}
In the following, we will re-arrange~\eqref{eqn:POD} so that we can determine the \ac{POD} modes and eigenvalues using an \ac{SVD} of a weighted snapshot matrix.
Inserting~\eqref{eqn:covarianceMatrix} into~\eqref{eqn:POD} yields
\begin{equation}
        \left(  \frac{\mathbf X \sqrt{\mathbf{W}}}{\sqrt{M-1}} \right)^H  \frac{\mathbf X \sqrt{\mathbf{W}}}{\sqrt{M-1}} \mathbf \Phi = \mathbf \Phi \mathbf \Lambda
\end{equation}
with $\sqrt{\mathbf W} = \diag\left(\sqrt{w_1}, \dots, \sqrt{w_N}\right)$.
The weighted snapshot matrix can be written as compact \ac{SVD}
\begin{equation} \label{eqn:SVD}
    \frac{\mathbf X\sqrt{\mathbf{W}}}{\sqrt{M-1}}  = \mathbf U \mathbf \Sigma \mathbf V^H, \quad
    \mathbf U \in \mathbb C^{R\times R}, \,
    \mathbf \Sigma \in \mathbb C^{R \times R}, \,
    \mathbf V \in \mathbb C^{N \times R}
\end{equation}
with $R = \min(M, N)$ and $\mathbf U^H\mathbf U = \mathbf V^H\mathbf V = \mathbf 1$.
This yields
\begin{equation}
    \mathbf V \mathbf \Sigma^H \mathbf \Sigma \mathbf V^H \mathbf \Phi = \mathbf \Phi \mathbf \Lambda.
\end{equation}
With~\eqref{eqn:POD_modes_orthonormal}, it can be written as
\begin{equation}
    \mathbf V \mathbf \Sigma^H \mathbf \Sigma \mathbf V^H = \sqrt{\mathbf W} \mathbf \Phi \mathbf \Lambda^H \left(\sqrt{\mathbf W} \mathbf \Phi\right)^H.
\end{equation}
So the eigenvalues $\mathbf{\Lambda}$ and modes $\mathbf{\Phi}$ are obtained from the weighted snapshot \ac{SVD} by
\begin{align} \label{eqn:SVD_to_POD_eigenvalues}
\mathbf \Lambda = \mathbf \Lambda^H &= \mathbf \Sigma^H \mathbf \Sigma
\\ \label{eqn:SVD_to_POD_modes}
\mathbf \Phi &= \sqrt{\mathbf W^{-1}} \mathbf V
\end{align}
with $\sqrt{\mathbf W^{-1}} = \diag\left(\sqrt{w_1^{-1}}, \dots, \sqrt{w_N^{-1}}\right)$.

\end{document}